## TITLE

# Lifshitz transitions and isospin polarization in twist-decoupled monolayer-bilayer graphene

Alex Boschi,[1] Leonardo Sabattini,[2,$] Sergey Slizovskiy,[3,4] Vaidotas Mišeikis,[1] Zewdu M. Gebeyehu,[1] Stiven Forti,[1] Antonio Rossi,[1] Kenji Watanabe,[5] Takashi Taniguchi,[6] Fabio Beltram,[2] Vladimir I. Fal'ko,[3,4,7] Camilla Coletti,[1] Sergio Pezzini,[2,†]

[1]Center for Nanotechnology Innovation @NEST, Istituto Italiano di Tecnologia, Piazza San Silvestro 12, I-56127 Pisa, Italy
[2]Istituto Nanoscienze – CNR, NEST - Scuola Normale Superiore, Piazza San Silvestro 12, 56127, Pisa, Italy
[3]National Graphene Institute, The University of Manchester, Manchester, M13 9PL, UK
[4]School of Physics & Astronomy, The University of Manchester, Oxford Rd., Manchester, M13 9PL, UK
[5]Research Center for Electronic and Optical Materials, National Institute for Materials Science, 1-1 Namiki, Tsukuba, 305-0044, Japan
[6]Research Center for Materials Nanoarchitectonics, National Institute for Materials Science, 1-1 Namiki, Tsukuba, 305-0044, Japan
[7]Henry Royce Institute for Advanced Materials, Manchester, M13 9PL, UK
[$]present affiliation: QuTech and Kavli Institute of Nanoscience, Delft University of Technology, Lorentzweg 1, 2628 CJ Delft, Netherlands

## ABSTRACT

Bernal-stacked bilayer graphene (BLG) hosts correlated electronic phases tied to low-energy Lifshitz transitions at saddle points in its valence band. To access this regime, ultralow charge disorder and control over a vertical electric field are simultaneously required. Here, we employ a twist-decoupled monolayer (MLG) to bias a proximal BLG in the absence of an external displacement field (*D*). We thereby reveal three-fold degenerate quantum Hall states at $D = 0$, with multiple transitions driven by doping, magnetic and electric field. Spontaneous broken symmetry in the vicinity of the valence band edge is signaled by the emergence of quantum oscillations with anomalous frequencies and large quasiparticle mass. These results indicate that electronic interactions in BLG are preserved in presence of an atomically close MLG, while showcasing the potential of CVD-grown graphene multilayers for the exploration of correlated phases of matter.

## I. INTRODUCTION

Crystalline graphene multilayers have become an increasingly popular platform for the investigation of electronic interactions in two dimensions. Thus far, reported correlated phases include spontaneous gaps [1–3], ferromagnets [4,5], Chern insulators [6], integer [7] and fractional [8,9] quantum anomalous Hall insulators, and superconductors [10], often connected with each other via electrostatic gating [11]. A key advantage of this class of systems is to bypass the issues related to structural disorder of moiré graphene [12], the archetypical correlated two-dimensional (2D) family [13]. However, trilayer and thicker crystalline graphene possess the required flat band features only in the case of rhombohedral (ABC) stacking, a metastable configuration that needs *ad hoc* strategies to avoid structural relaxation [4,14,15]. Bernal-stacked (AB) bilayer graphene (BLG) samples, conversely, can be handled via conventional van der Waals techniques [16], while displaying exotic properties analogous to their thicker counterparts [17–22]. Nonetheless, the large density of states in BLG is confined to a narrow energy range (~1 meV, approximately a factor of ten smaller than in ABC trilayer [23]), implying that devices with ultralow electrostatic disorder are necessary [24]. In addition, control over a vertical displacement field (*D*) is needed to enhance saddle-point van Hove singularities (vHs) [25] associated to single-particle gap opening induced by layer polarization [26,27]. A different route to gapping BLG in absence of a *D* field is offered by large-angle twisting either with another BLG [28] or a graphene monolayer (MLG) [29,30]. This approach takes advantage of electronic decoupling due to momentum mismatch [31–33], combined with proximity-induced energy shifts resulting in layer-polarized BLG subsystems. An energy shift of 18 meV was recently measured in twisted monolayer-bilayer graphene (TMBG) encapsulated in hexagonal boron nitride (hBN) [29], corresponding to the effect of a sizeable external $D$ field (~ 0.15 V/nm, as confirmed by Jiang et al. [30]). Therefore, TMBG has the potential to offer $D$ field-free access to the valence band of biased BLG. In addition, proximal graphene layers result in considerable screening effects [34], which might influence the correlated phase diagram with respect to that of free standing BLG.

[†]Contact author: sergio.pezzini@nano.cnr.it

In this Letter, we present low-temperature (magneto)transport experiments on a low-disorder TMBG dual-gated device, in which the built-in asymmetry gaps the BLG subsystem. We observe quantum oscillations associated with a three-fold valence band edge at $D$ = 0. The evolution of the oscillations under different external fields indicates tunable Lifshitz transitions (LTs) [35] across a saddle-point vHs, consistent with the single-particle band structure of biased BLG. We further reveal oscillations beyond the non-interacting picture, signaling broken-symmetry phases stabilized by Coulomb interaction. We discuss these findings in the context of previous studies on free standing BLG. Our results establish CVD-grown large-angle twisted graphene as a platform for many-body electron physics.

## II. RESULTS AND DISCUSSION

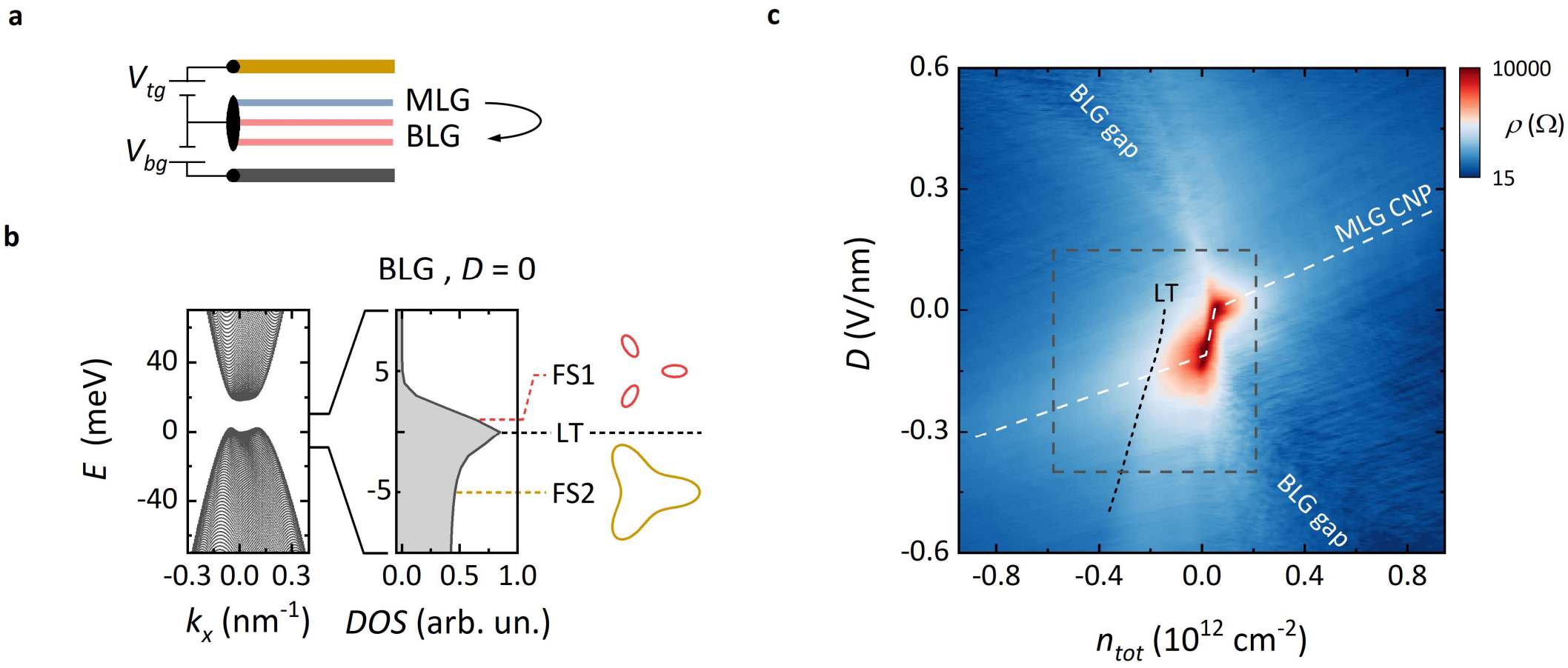

FIG. 1: *Accessing the valence band of biased BLG within TMBG. (a) Sketch of the studied device with gating configuration. Yellow: Au gate; dark gray: graphite gate; blue: MLG; red: BLG. (b) Left: low-energy BLG bands* within TMBG *at* $D$ = 0 *(left), as a function of* $k_x$ *relative to the K point. Center: density of states in the top part of the BLG valence band (-10 meV <* $E$ *< +10 meV). Right: evolution of the Fermi surface across the LT. FS1 (red) is shown at* $E$ *= 1 meV, FS2 (yellow) is shown at* $E$ *= -5 meV. (c) Resistivity as a function of gate-induced charge density and displacement field, measured at* $T$ *= 0.4 K and* $B$ *= 0 T. A log color scale is used. The black dashed line is the calculated n-D position of the BLG LT. The white dashed line is a guide to the eye for the MLG CNP. The dashed gray rectangle indicates the n-D ranges adopted in the following parts of the manuscript.*

Figure 1a shows the device configuration. An hBN-encapsulated CVD-grown TMBG crystal, composed of a MLG (blue) twisted by 30° on top of BLG (red), is controlled by a top (Au, yellow) and a bottom (graphite, dark gray) gate electrode. The combined action of the gate voltages ($V_{bg}$ and $V_{tg}$) induces a total charge density $n_{tot} = 1/e \cdot (C_{bg}V_{bg} + C_{tg}V_{tg})$ and a displacement field $D/\varepsilon_0 = (C_{bg}V_{bg} - C_{tg}V_{tg})/2$ [31], where $C_{bg}$ and $C_{tg}$ are the capacitances per unit-area of the two gates. The sample, fabricated according to the methods reported in Ref. [29] and referred to as device D2 therein, is the TMBG with lowest disorder among those at disposal at the time of performing this experiment. The presence of MLG results in a band gap in the BLG subsystem even at $D = 0$, as shown in the calculations of Figure 1b, left (details on the model are reported in Supplemental Material (SM) [36]). Close to the valence band maximum, a saddle point is visible, leading to a vHs in the density of states (DOS, Figure 1b, center). The BLG vHs is associated with a ferromagnetic instability [25], which drives a cascade of isospin (valley and spin) phase transitions in free standing BLG [20,23,37]. In a single-particle picture [38,39], the vHs is signaled by a characteristic change in the topology of Fermi surface (FS), that is a LT [35]. As shown in Figure 1b, right, at larger hole doping, the topology of the Fermi surface evolves from being three disconnected pockets (FS1) into a single hole pocket (FS2), each being four fold degenerate in spin and valley.
Resistivity ($\rho$) measurements at low temperature ($T = 0.4$ K) and zero magnetic field, presented in Figure 1c, give a first indication of whether this scenario and the associated many-body physics may be detected in our CVD-grown TMBG. Based on previous works [29,30], we identify the feature at highest $\rho$ (dark red color) as the MLG charge neutrality point (CNP) crossing the BLG band gap. Outside of the BLG gap, the MLG CNP acquires a smaller slope due to screening by the BLG DOS, following the trajectory depicted in Figure 1c (white dashed line), and confirmed by measurements of the MLG $N = 0$ Landau level (0-LL) at finite magnetic field (discussed later). The BLG gap collapses at $D \sim 0.15$ V/nm [29,30]. On the left side of the gap (i.e., in the BLG valence band) we observe a region with moderate $\rho$ (light red to white color) that fans out of the gap closing point and expands toward negative displacement field. $\rho$ shows a steep increase upon entering this region (see Figure S1) – which aligns with the position of the BLG LT obtained from our TMBG model (dashed black line in Figure 1c) – and a series of peaks reminiscent of free standing BLG under comparable displacement fields [20,37] (see Figure S2). Such features, previously unreported in large-angle TMBG, define the *n-D* region where we focus for the remainder of our study (dashed gray rectangle in Figure 1c).

We now address the response of the device to a finite magnetic field. At $D = 0$, we measure the LL fan shown in Figure 2a. The contribution of MLG LLs is restricted to the high-conductivity region (red color) at low magnetic field ($B < 0.3$ T), while MLG keeps at $\nu_{MLG} = -2$ throughout the majority of the map. The pinning of $\nu_{MLG}$ is evidenced by the straight trajectories of BLG LLs, following $n_{tot}\, h/eB$. We can thus focus on quantum Hall (QH) features deriving from the BLG valence band, with associated filling factor and Hall conductivity quantified as $\nu_{BLG} = \nu_{tot} - \nu_{MLG}$ and $\sigma^{BLG}_{xy} = \sigma_{xy} - \sigma^{MLG}_{xy}$, respectively. The $\nu_{BLG} = 0$ QH state extends to vanishing $B$ due to the gap of layer-polarized BLG. In the valence band, we observe different QH sequences. At moderate magnetic field (see red dashed line in Figure 2a) we observe three-fold degenerate QH states ($\nu_{BLG} = -3, -6$), which evolve into four fold degenerate ones upon hole doping. Line cuts of longitudinal conductivity ($\sigma_{xx}$) and $\sigma^{BLG}_{xy}$, shown in Figure 2b, confirm the change in degeneracy, which marks the underlying doping-induced LT from FS1 to FS2 at zero field. According to single-particle calculations [38,39] FS1 should give a first QH state at $\nu_{BLG} = -6$, due to the presence of three pockets and spin degeneracy (the valley degeneracy is lifted in the LL spectrum of gapped BLG). $\nu_{BLG} = -3$ is stabilized by exchange interaction and shows comparable robustness to $\nu_{BLG} = -6$, as also reported for BLG in Ref. [40]. By increasing the magnetic field, we observe a LL crossing (white arrow in Figure 2a) that alters the QH sequence. As shown in Figure 2c, above this LL crossing we observe all integer QH states, with even (spin-unpolarized) states more robust than odd (spin-polarized) ones. Hence, irrespective of the doping level, QH states characteristic of a single hole pocket are observed (see sketch in Figure 2c). This transition was first identified in Ref. [38] and attributed to electron like LLs in the BLG valence band that do not mix with hole-like ones. We note that the same change in the QH sequence could be also driven by magnetic breakdown [40], considering that the *k*-space separation of the pockets in FS1 is comparable to the inverse of the magnetic length ($\sim 0.06$ nm$^{-1}$ at $B = 2$ T).

The field-driven QH transitions observed at $D = 0$ can be further tuned by the $D$ field, as exemplified by data at $B = 1.2$ T shown in Figure 2d. In this map, BLG QH states disperse vertically, unless shifting by $\nu_{tot} = +4$ when crossing the MLG 0-LL. We detect again a transition close to the valence band edge, positioned at $D \sim 0$: for $D < 0$ (red dashed line) we observe three-fold QH states, while only two-fold states are visible at $D > 0$ (yellow line).

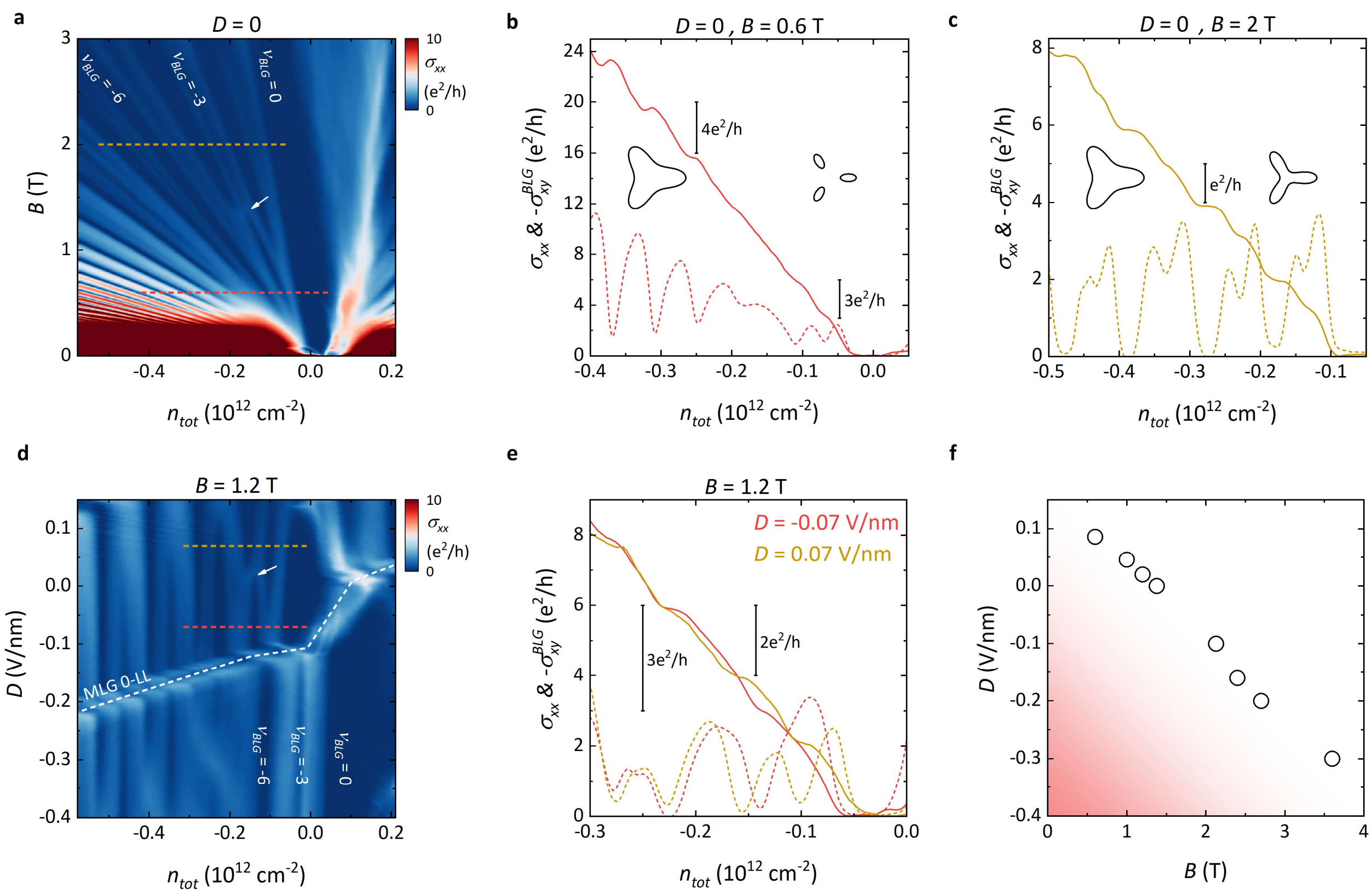

FIG. 2: *Tunable BLG Lifshitz transitions in TMBG. (a) Longitudinal conductivity as a function of magnetic field and carrier density, measured at D = 0. The red and yellow dashed lines locate data plotted in panel b and c, respectively. The white arrow indicates a LL crossing close to the BLG valence band edge. (b) Longitudinal (dashed line) and Hall (continuous line) conductivity, measured along the red dashed line in panel a (B = 0.6 T). The evolution of the FS induced by hole doping is sketched. (c) Same as panel b, measured along the yellow dashed line in panel a (B = 2 T). (d) Longitudinal conductivity as a function of displacement field and carrier density, measured at B = 1.2 T. The red and yellow dashed lines locate data plotted with the same colors in panel e. The white arrow indicates the same LL crossing individuated in panel a. (e) Longitudinal (dashed line) and Hall (continuous line) conductivity, measured along the red and yellow dashed lines in panel d (D = - 0.07 V/nm and D = 0.07 V/nm, respectively). (f) Position of the LL crossings as a function of magnetic and displacement field. FS1 is observable only on the lower-left side of the diagram (indicated by the red shading). For better visibility,* $\sigma_{xx}$ *traces are multiplied by a factor ×2, ×10, ×4 in panels b, c, e, respectively.*

The corresponding conductivity line traces are shown in Figure 2e. The phenomenology is similar to the $B$-driven transition discussed above, and consistent with previous observations in biased free standing BLG [38–40]. We note that odd states are absent for $D > 0$, while $\nu_{BLG} = -3$ is present for $D < 0$ at the same magnetic field, suggesting a different stabilization mechanism of spin-polarized states on the two sides of the of the QH transition. A $B$-$D$ phase diagram is shown in Figure 2f, where we plot the position of the transition from additional measurements either at fixed $B$ or $D$ (shown in Figure S3 and Figure S4, respectively). The magnetic field required for the transition increases with |D| for D < 0, while it reduces for D > 0, showing a trend toward vanishing at the gap closing point ($D$ = 0.15 V/nm, where we measure a conventional four fold LL fan, shown in Figure S4).

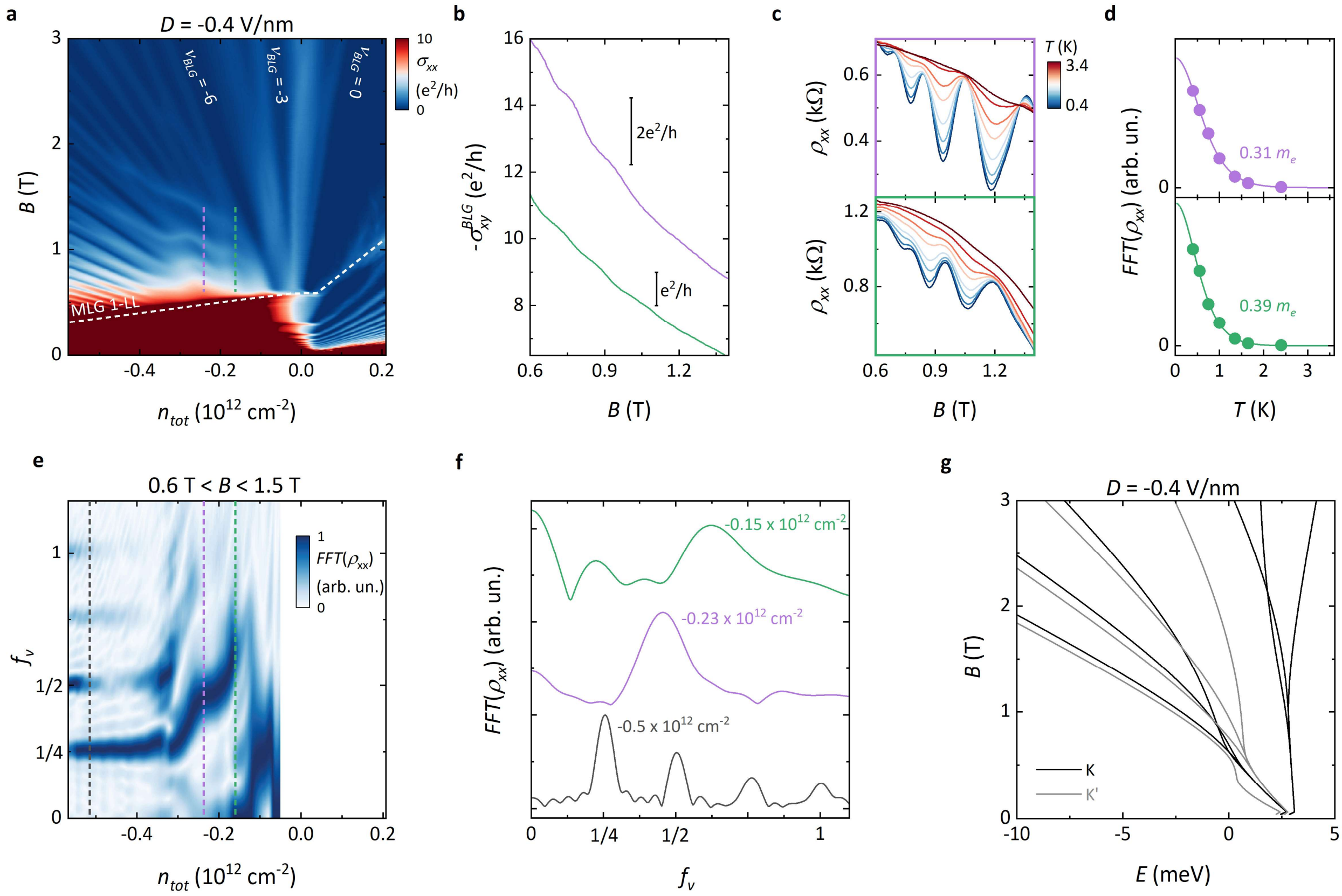

FIG. 3: *Signatures of spontaneous isospin polarization of BLG holes. (a) Longitudinal conductivity as a function of magnetic field and carrier density, measured at D = -0.4 V/nm. The pink and green dashed lines identify data plotted in panels b and c. (b) Hall conductivity measured along the dashed lines in panel a. (c) Longitudinal resistivity at selected temperatures, measured along the dashed pink (upper) and green (lower) lines in panel a. (d) Amplitude of the FFT peak as a function of temperature (dots), corresponding to the resistivity oscillations in panel c. The continuous lines are fits to the data using the Lifshitz–Kosevich formula. (e) FFT power spectral density as a function of carrier density and frequency, calculated from data in panel a. The dashed lines correspond to individual spectra plotted in panel f. (f) FFT spectra at selected carrier density. (g) Single-particle calculations of BLG LLs in biased TMBG. Black (gray) lines correspond to levels from the K (K') valley. Each level has a two-fold spin degeneracy.*

We next focus on the behavior at large negative displacement field ($D$ = -0.4 V/nm), where the vHs is enhanced, shifts toward higher hole density, and measurements in magnetic field are less restricted by the transitions discussed in the previous section. Comparison with analogous data from free standing BLG [20,23,37] can be made by considering the $D$ field acting on BLG as the sum of the gate induced one (see Figure S2) and the one due to energy shifts in hBN-encapsulated TMBG, resulting in ~ -0.55 V/nm. Figure 3a shows a rich LL fan diagram. MLG electron LLs are progressively filled in the lower part of the plot (up to ~0.6 T), implying that $\nu_{MLG}$ = +2 should be considered at higher fields. Therefore, the slope of BLG QH states shifts by a factor +4 with respect to the fan diagram measured at $D = 0$ (Figure 2a). Close to the valence band edge, QH states at $\nu_{BLG} = 0, -3, -6$ can be again identified. $\nu_{BLG} = -3$ extends uninterrupted over the whole magnetic field range considered, as expected from the positioning on the phase diagram in Figure 2f. Nonetheless, we observe a series of crossings at partial integer fillings ($\nu_{BLG} = -1, -2$ and $\nu_{BLG} = -4, -5$, see also Figure S5). According to Ref. [37] these structures arise from LLs crossings for the non-degenerate case, thus indicating isospin polarization in FS1. Further crossings and non-monotonous behavior are visible at higher hole density, qualitatively similar to observations in Ref. [37], eventually leading to the standard four fold QH states associated to FS2. We focus on the intermediate doping range ($-0.3 \times 10^{12}$ cm$^{-2}$ < $n_{tot}$ < $-0.1 \times 10^{12}$ cm$^{-2}$) in the vicinity of the vHs, where we identify two separate sets of oscillations (pink and green dashed lines in Figure 3a) which were not detected at $D = 0$. The corresponding Hall conductivities (Figure 3b) show approximately double and single quantized steps, consistent with an approximate double and single spacing in filling factor (see Figure S5), respectively. The oscillations are quickly suppressed by temperatures of a few Kelvin (Figure 3c), following a Lifshitz–Kosevich-type dependence [41] (see fits in Figure 3d and SM for analysis details [36]). The extracted effective mass $m^*$ is found to be within 0.3 - 0.4 electron mass ($m_e$) for both oscillations. Such values, one order of magnitude larger with respect to measurements on unbiased BLG at similar carrier density [42], are indicative of band flattening at the vHs. Following Refs. [20,23], we consider the fast Fourier transform (FFT) of resistivity data as a function of $1/B$ (see Supplemental Material for technical details [36]), presented as a color map as a function of $n_{tot}$ in Figure 3e (selected line traces are shown in Figure 3f). We limit this analysis to 0.6 T < $B$ < 1.5 T, where the two high-mass oscillations previously discussed are clearly discernible, and avoiding contributions from the MLG LLs. On the y-axis, we divide the oscillation frequency $f_B$ by the carrier density and the magnetic flux, obtaining $f_\nu = f_B / (n_{tot}\, h/e)$, which corresponds to the fraction of carriers enclosed by the coherent orbit producing the $1/B$-periodic resistivity oscillation [4]. At large hole densities (dark gray dashed line), we observe a main peak at $f_\nu = 1/4$, indicating four Fermi contours with equal carrier population, as expected for FS2. This FFT peak progressively shifts to $f_\nu > 1/4$ with reducing hole density, consistent with the nucleation of a small electron orbit [4]. Close to $n_{tot} \sim -0.3 \times 10^{12}$ cm$^{-2}$ the FFT spectrum shows discontinuity, followed by a progressive increase in the frequency of the main peak, which saturates close to $f_\nu = 1/2$. At this point, we encounter the first high-mass oscillation described above (pink dashed line), consistent with the approximate 2-fold steps in $\sigma^{BLG}_{xy}$ (Figure 3b). However, the fact that the peak frequency is always $f_\nu <$ 1/2 indicates the presence of a population of minority carriers with a different spin-valley flavor, in line with partially isospin polarized (PIP) phases previously identified in BLG [20,23]. The same picture of partial isospin polarization applies to the second oscillation (green dashed line), which shows a frequency $1/2 < f_\nu < 1$. Here, we observe also lower frequency components in the FFT, which might arise from the minority carriers.

Single-particle calculations for BLG under a comparable $D$ field, shown in Figure 3g, predict lifting of the valley degeneracy and a series of LLs crossings. However, the salient features identified experimentally, i.e., partial isospin polarization and the two high-mass oscillations at intermediate densities, are not expected from the BLG LL spectrum (an interacting band model would be likely required, see for instance Ref. [43]). Despite the proximity of MLG, which has been shown to effectively screen disorder in large-angle twisted bilayer graphene [34], the observed sequence of PIP phases agrees with those reported in Refs. [20,23]. As such, the screening effect of MLG does not appear to suppress or significantly alter the phase diagram of gapped BLG. Our results are in accordance with experiments on magic-angle graphene with a proximal large-angle twisted MLG [44], which show preservation of correlated states from the flat-band subsystem.

## III. CONCLUSIONS

In conclusion, we showed that twist-decoupled monolayer-bilayer graphene stack offers a clean and controllable way to access the regime where Bernal bilayer graphene develops strong interaction effects. We uncover tunable changes in Fermi-surface topology and high-mass oscillations in correspondence of partially isospin-polarized states – showing that key features of the correlated physics of BLG are preserved in the presence of a nearby MLG. These results establish CVD-based TMBG crystal as a scalable platform capable of hosting the rich many-body behavior previously accessible only in the cleanest exfoliated devices, broadening the experimental landscape for studying correlated electrons in multilayer graphene and opening new opportunities for engineered van der Waals systems.

## ACKNOWLEDGMENTS

A.B., V.M. and C.C. acknowledge financial support from the European Union through the GraPh-X project, under the grant agreement 101070482. F.B., C.C. and S.P. acknowledge financial support from PNRR MUR project No. PE00000023 - NQSTI. K.W. and T.T. were supported by the JSPS KAKENHI (Grant Numbers 21H05233 and 23H02052), the CREST (JPMJCR24A5), JST and World Premier International Research Center Initiative (WPI), MEXT, Japan, for hexagonal boron nitride crystals growth. V.F. and S.S. acknowledge support from EPSRC grant EP/V007033/1, British Council and International Science Partnerships Fund Grant 1185409051 for Research Collaboration between UK and Japan.

## DATA AVAILABILITY

The data that support the findings of this Letter are openly available [45].

---

**SUPPLEMENTAL MATERIAL FOR**

**Lifshitz transitions and isospin polarization in twist-decoupled monolayer-bilayer graphene**

Alex Boschi,[1] Leonardo Sabattini,[2,$] Sergey Slizovskiy,[3,4] Vaidotas Mišeikis,[1] Zewdu M. Gebeyehu,[1] Stiven Forti,[1] Antonio Rossi,[1] Kenji Watanabe,[5] Takashi Taniguchi,[6] Fabio Beltram,[2] Vladimir I. Fal'ko,[3,4,7] Camilla Coletti,[1] Sergio Pezzini,[2,†]

[1]Center for Nanotechnology Innovation @NEST, Istituto Italiano di Tecnologia, Piazza San Silvestro 12, I-56127 Pisa, Italy
[2] Istituto Nanoscienze – CNR, NEST - Scuola Normale Superiore, Piazza San Silvestro 12, 56127, Pisa, Italy
[3]National Graphene Institute, The University of Manchester, Manchester, M13 9PL, UK
[4]School of Physics & Astronomy, The University of Manchester, Oxford Rd., Manchester, M13 9PL, UK
[5]Research Center for Electronic and Optical Materials, National Institute for Materials Science, 1-1 Namiki, Tsukuba, 305-0044, Japan
[6]Research Center for Materials Nanoarchitectonics, National Institute for Materials Science, 1-1 Namiki, Tsukuba, 305-0044, Japan
[7]Henry Royce Institute for Advanced Materials, Manchester, M13 9PL, UK
[$]current affiliation: QuTech and Kavli Institute of Nanoscience, Delft University of Technology, Lorentzweg 1, 2628 CJ Delft, Netherlands

**Supplemental Note 1. Band structure calculations of BLG within TMBG**

We describe the BLG in terms of effective k·p Hamiltonian in the vicinity of the respective K-points as

$$H_{BLG} = \begin{pmatrix} U_1 + \delta & v\,\pi^\dagger & -v_4\,\pi^\dagger & v_3\pi \\ v\,\pi & U_1 + \delta + \Delta_{AB} & \gamma_1 & -v_4\pi^\dagger \\ -v_4\pi & \gamma_1 & U_2 + \Delta_{AB} & v\,\pi^\dagger \\ v_3\pi^\dagger & -v_4\pi & v\,\pi & U_2 \end{pmatrix}.$$

Here, the operators $\pi$ are complex momenta, $\mp i\,\hbar\,\partial_x + \hbar\,\partial_y - i\,e\,B\,x,$ counted from $K_\pm$ points of BLG; $v = 1.02 \cdot 10^6$ m/s is the Dirac velocity in BLG, parameters $v_i = \frac{\sqrt{3}}{2} a\,\gamma_i$ are determined by Slonczewski-Weiss-McClure skew-hopping energies, where $a = 0.246$ nm is graphene's lattice constant. We use the standard values for trigonal warping $\gamma_3$ =0.38 eV, vertical interlayer hopping $\gamma_1$ =0.39 eV and dimer—non-dimer energy difference $\Delta_{AB} = 0.02$ eV in BLG, while we use the enlarged value of dimer--non-dimer hopping $\gamma_4$ =0.3 eV to match the location of hole-side Lifshitz transition. The interlayer-proximity-induced energy shift $\delta = 18$ meV between layers 1 and 3 which interface the encapsulating hBN and the middle layer 2 was found in previous publication [1].

**Supplemental Note 2. Resistivity curves at zero magnetic field**

In Figure S1 we show three line-cuts from resistivity data in main text Figure 1c. In the measurement configuration adopted, BLG and MLG act as parallel conducting channels. Although it is difficult to quantify the resistivity of the individual subsystem, we can assign several features to each of them. Specifically, we ascribe to BLG the clear increase in resistivity at $n_{tot}$ = -0.4 × $10^{12}$ cm$^{-2}$, as well as the fluctuations at lower carrier density. Previous studies on BLG locate the vHs in this density region at similar displacement field [2–4]. Transport measurements therein also report resistivity peaks in this region [2,3], as also visible also in our data. Temperature dependent measurements at large displacement field are shown in Figure S1b. Starting from the BLG band edge (located at $n_{tot}$ ~ 0.1 × $10^{12}$ cm$^{-2}$), we progressively observe one insulating region, followed by two separated metallic ones.

Similar alternating insulating-metallic pockets are reported in Ref. [3]. A quantitative comparison is, however, beyond our possibilities due to the MLG-BLG parallel conducting configuration.

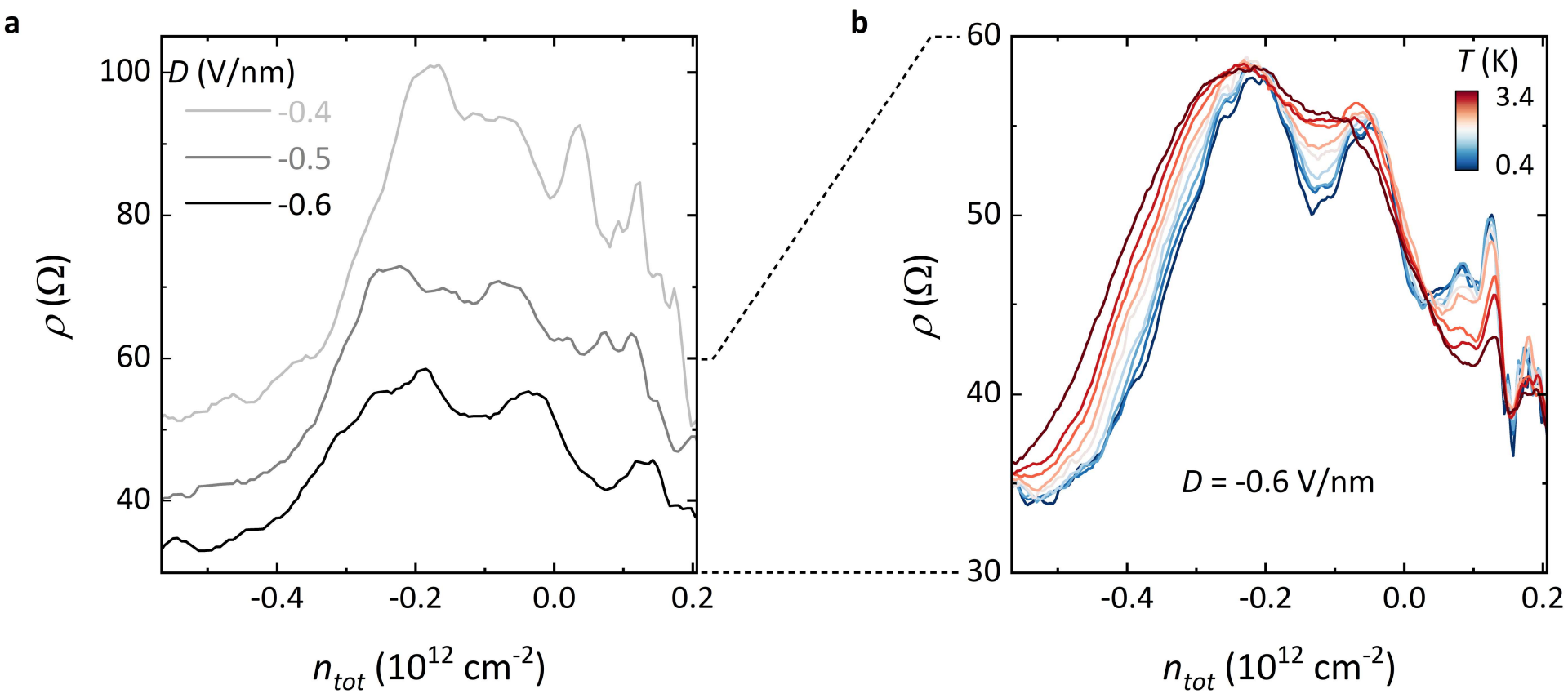

FIG. S1. *(a) Resistivity as a function of carrier density at selected displacement fields. (b) Resistivity as a function of carrier density at D = -0.6 V/nm, measured at different temperatures.*

**Supplemental Note 3. Displacement field experienced by BLG**

To compare literature data on BLG with our TMBG system, one needs to refer to the displacement field acting on BLG, rather than the value induced by gates. However, once the energy shifts on the outer layers are considered, the difference between the two displacement fields turns out to be practically negligible. Following the notation of the self-consistent model presented in Ref. [1] (shown below in Figure S2a), the displacement field on the whole stack is given by $D_{Gate} = (D_0 + D_3)/2$: this is the quantity that we control experimentally via the gates, and we refer to it as $D$ throughout the paper. The displacement applied to BLG is given by $D_{BLG} = (D_0 + D_2)/2$. The difference between the two fields stems from the electric field of MLG and it is proportional to MLG doping. This contribution is typically very small, due to the different quantum capacitance of the two subsystems. In panel (Figure S2b) below, we show a plot of the energy gap induced on BLG as a function of $D_{Gate}$ (full lines) or $D_{BLG}$ (dashed lines), in the case with (dark cyan) and without (black) energy shifts (δ).

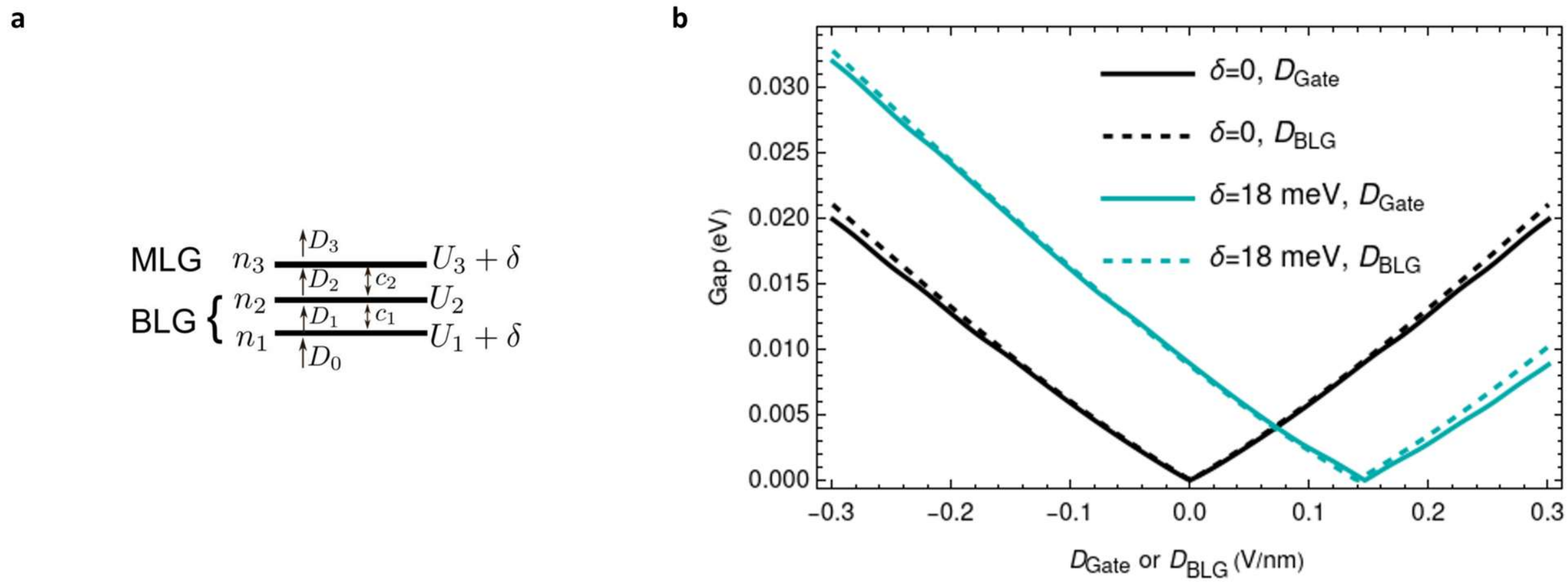

FIG. S2 *(a) Sketch of the TMBG system used for the self-consistent model described in Ref.* [1]. *(b) Band gap of BLG within TMBG, as a function of the gate induced D (continuos lines) and D experienced by BLG (dashed*

*lines). The black (dark cyan) lines show the results of calculations not considering (including) the energy shifts due to the structural asymmetry of the hBN-encapsulated TMBG.*

**Supplemental Note 4. Additional conductivity maps at finite magnetic field**

Figure S3 presents conductivity maps as a function of carrier density and displacement field (analogous to Figure 2d in the main text), measured at different magnetic fields. The white arrow in panels b, c, d and e indicates the LL crossing described in the main text. Figure S4 shows conductivity maps as a function of carrier density and magnetic field (analogous to Figure 2a and 3a in the main text), acquired at different displacement fields. The white arrow in panels b, c and d indicates again this LL crossing. The phase diagram in main text Figure 2f is constructed based on these measurements.

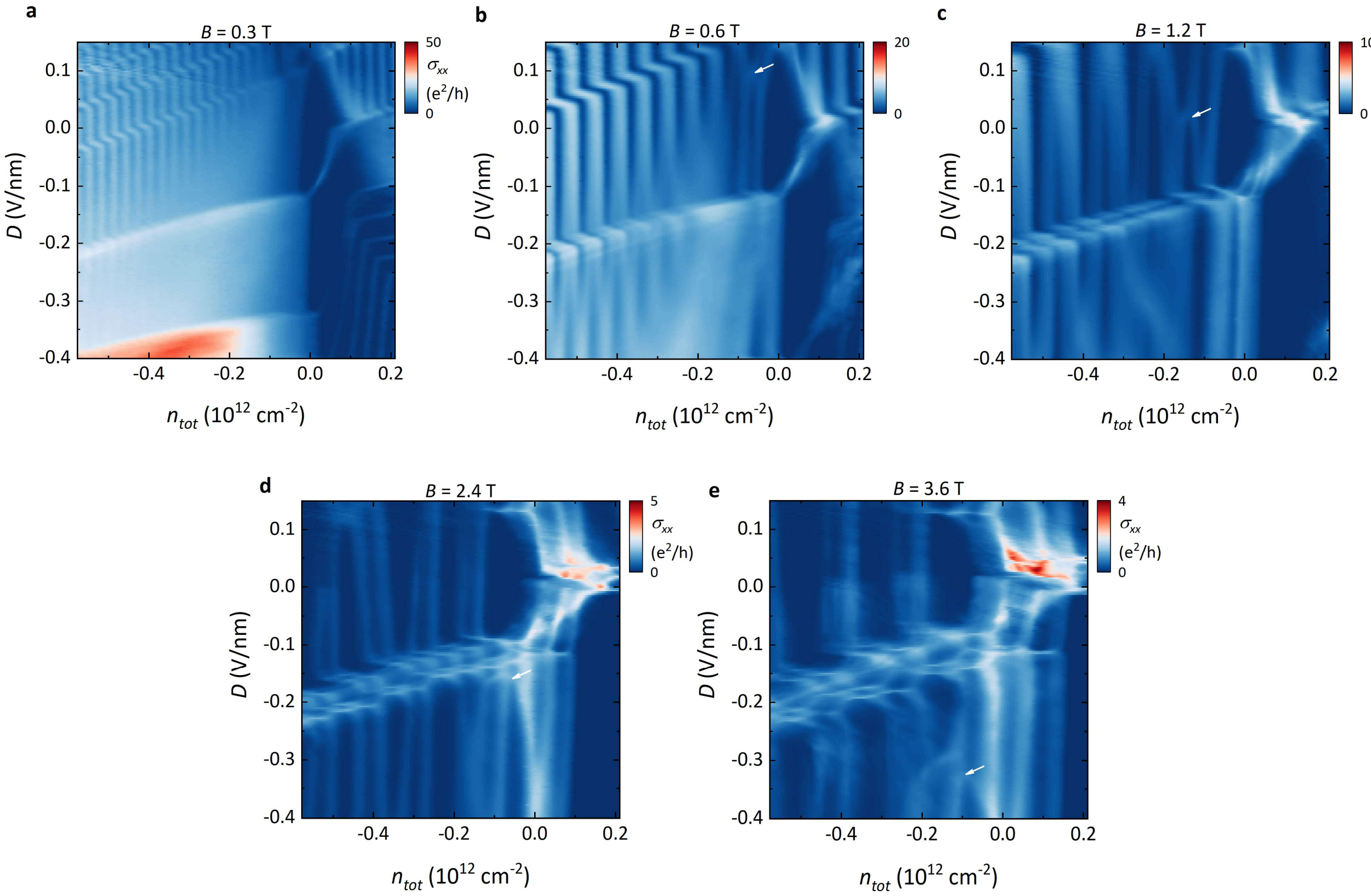

FIG. S3. *Longitudinal conductivity as a function of carrier density and displacement field, measured at different magnetic fields.*

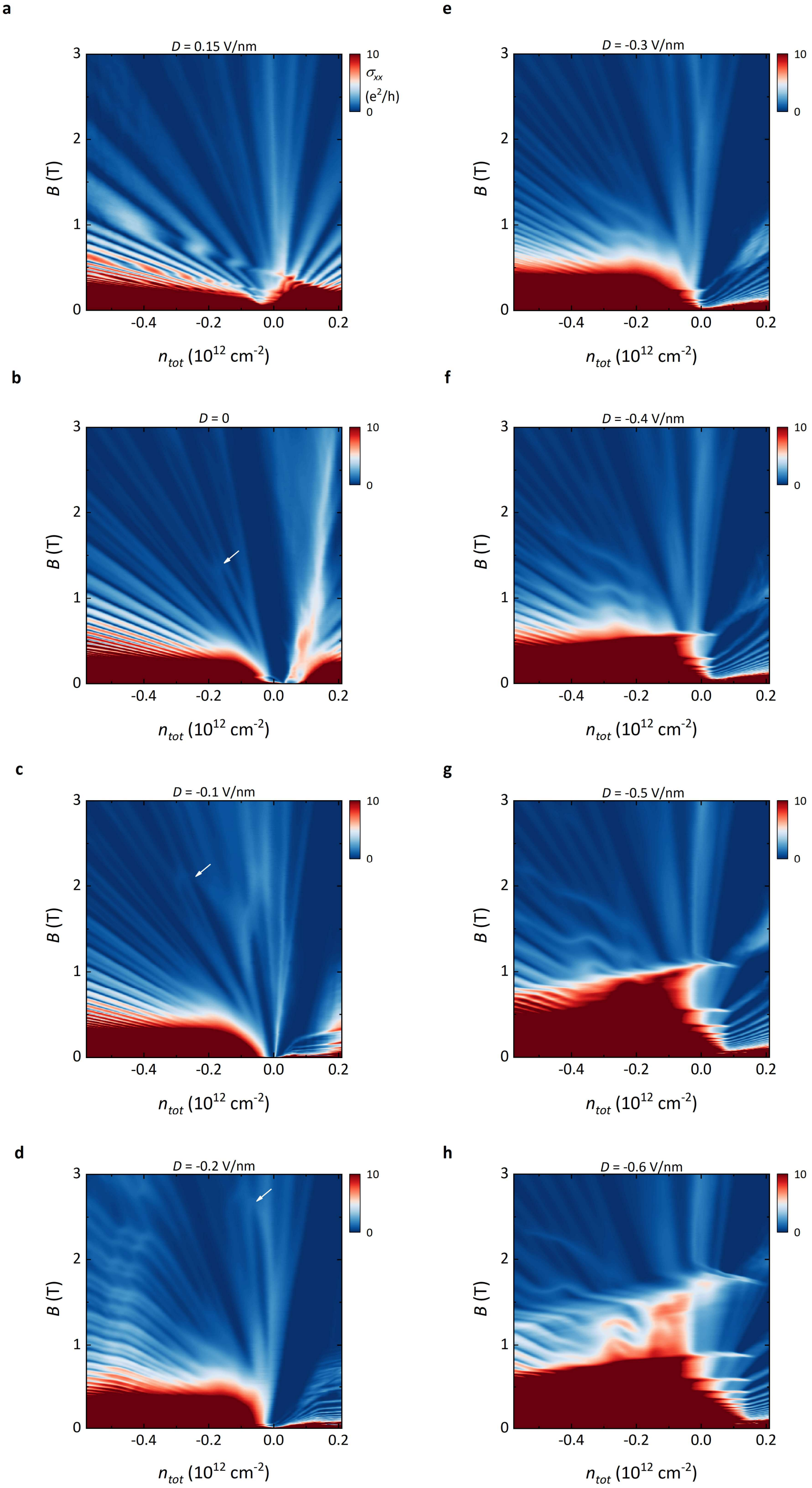

FIG. S4. *Longitudinal conductivity as a function of carrier density and magnetic field, acquired at different displacement fields.*

**Supplemental Note 5. Zooms on LL crossings and high-mass oscillations at *D* = -0.4 V/nm**

Figure S5 highlights three areas of Figure 3a, showing the numerical derivative of the longitudinal conductivity as a function of carrier density ($d\sigma_{xx}/dn_{tot}$). The derivative is employed to enhance the visibility of oscillatory features. On top of the data, we show the dispersion of several filling factors (colored dashed lines) calculated as $\nu_{BLG} = \nu_{tot} - \nu_{MLG}$ where $\nu_{MLG} = +2$. In panel a, we focus on the valence band edge. The red arrows indicate crossings at $\nu_{BLG} = -4$ and $\nu_{BLG} = -2$ consistent with the evolution of the LLs in the non-degenerate case [3]. In panel b and c we zoom on the two high-mass oscillations analyzed in the main text. The indicated filling factors are consistent with the quantized steps of $\sigma_{xy}$ shown in Figure 3b. The oscillations approximately follow the trajectories of such fillings.

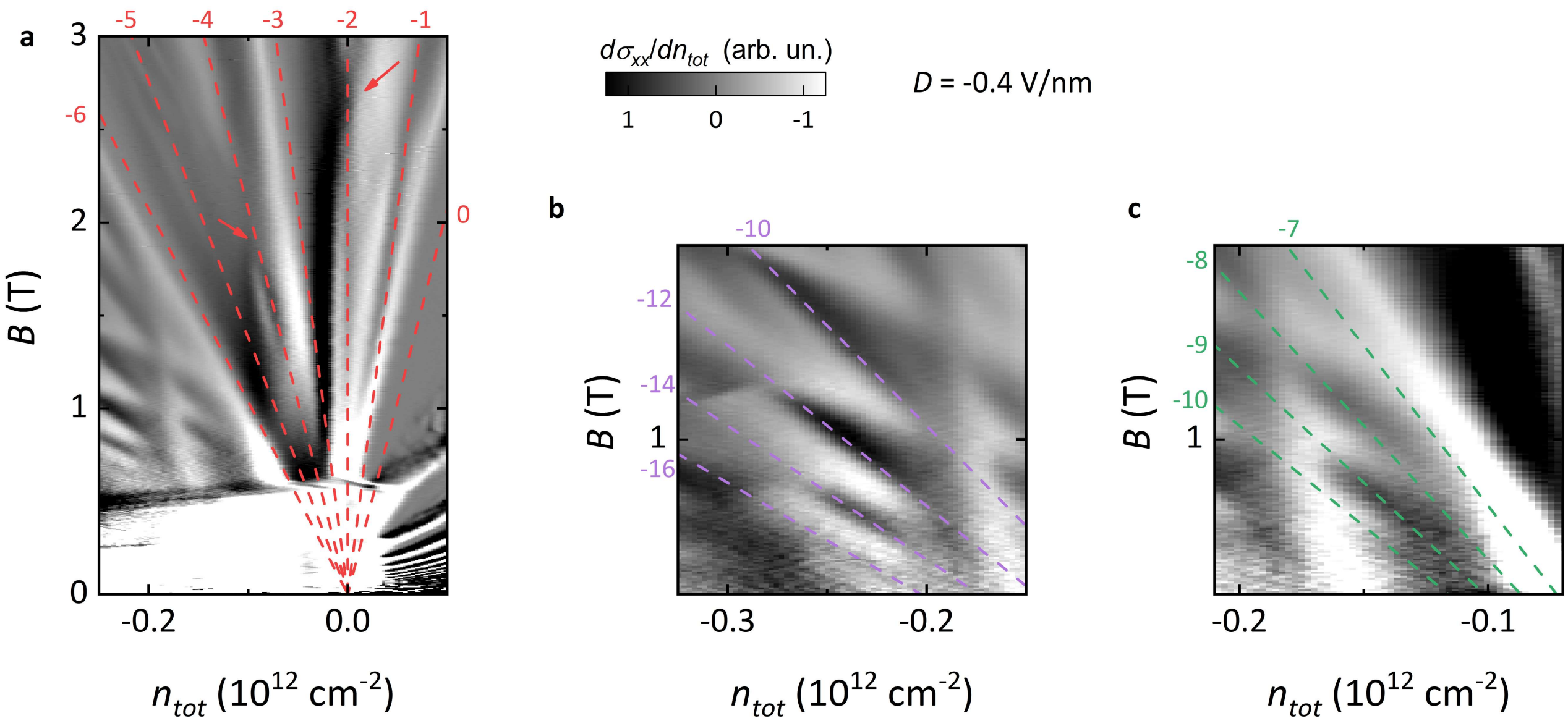

FIG. S5. *Numerical derivative of the longitudinal conductivity, as a function of carrier density and magnetic field, measured over different ranges. (a) Dashed red lines track all the integer filling factors observable close to the BLG band edge. The red arrows indicate crossings at* $\nu_{BLG} = -4$ *and* $\nu_{BLG} = -2$. *(b) Dashed pink lines track even filling factors in the region of the first high-mass oscillation. (c) Dashed green lines track all integer filling factors in the region of the second high-mass oscillation.*

**Supplemental Note 6. Methods for extraction of the effective mass and FFT**

To extract the effective mass $m^*$, we collect isotherms of $\rho_{xx}$ in correspondence of the two oscillations indicated by pink and green dashed lines in Figure 3a. These measurements are presented in Figure 3c. For each isotherm, the FFT is computed following the procedure detailed in the next paragraph. For each temperature, we then consider the amplitude of the FFT peak corresponding to the resistivity oscillation, FFT($\rho_{xx}$). The temperature-dependence of FFT($\rho_{xx}$) is fitted to the thermal damping factor of the Lifshitz–Kosevich formula,

$$\mathrm{FFT}(\rho_{xx})\,(T) = A_0 \cdot (\chi / \sinh(\chi)),$$

where $\chi = (2\pi^2 \cdot k_B \cdot m^* \cdot T) / (B_{eff} \cdot \hbar \cdot e)$, with $k_B$ the Boltzmann constant, $\hbar$ the reduced Planck constant and e the electron charge. $A_0$ is a temperature-independent pre-factor. The magnetic field $B_{eff}$ is taken as the inverse of the center of the $1/B$ window used for the FFT, i.e. $B_{eff} = 1/[(1/B_{max} + 1/B_{min})/2]$. Considering the standard error from the fits shown in Figure 3d, we obtain the following values and uncertainties for the two oscillations: $(0.310 \pm 0.004) \cdot m_e$ and $(0.390 \pm 0.010) \cdot m_e$.

The FFT is computed using data in the range $0.63\ \mathrm{T} < B < 1.5\ \mathrm{T}$ to exclude the monolayer graphene contribution; data outside this range are set to zero (zero padding). The data points are equally spaced in $B$, hence they are unevenly spaced in 1/B. Therefore, the data are first interpolated onto an equidistant grid, as required by FFT. A Welch window is applied prior to the transform to reduce spectral leakage. For clarity, in Figure 3e, each power spectral density associated with a given carrier density is normalized to its maximum.

---